\documentclass{moriond}
\usepackage{siunitx}
\def\be{\begin{equation}}
\def\ee{\end{equation}}
\def\bea{\begin{eqnarray}}
\def\eea{\end{eqnarray}}

\usepackage{hepunits}
\usepackage{amsmath}
\usepackage{xspace}

\newcommand{\PQt}{\ensuremath{t}\xspace}
\newcommand{\PAQt}{\ensuremath{\bar{t}}\xspace}
\newcommand{\etat}{\ensuremath{\eta_{t}}\xspace}
\newcommand{\chit}{\ensuremath{\chi_{t}}\xspace}

\newcommand{\ttbar}{{\PQt{}\PAQt}\xspace}
\newcommand{\mttbar}{\ensuremath{m_{\ttbar}}\xspace}

\newcommand{\chel}{\ensuremath{c_\text{hel}}\xspace}
\newcommand{\chan}{\ensuremath{c_\text{han}}\xspace}
\newcommand{\hvq}{\ensuremath{\texttt{hvq}}\xspace}
\newcommand{\POWHEG} {{\textsc{powheg}}\xspace}
\newcommand{\PYTHIA} {{\textsc{pythia}}\xspace}
\newcommand{\HERWIG} {{\textsc{herwig}}\xspace}
\newcommand{\POWHEGvtwohvqPYTHIA}{{\POWHEG v2 \hvq}\,+\,{\PYTHIA}\xspace}

\newcommand{\POWHEGhvqPYTHIA}{{\POWHEG \hvq}\,+\,{\PYTHIA}\xspace}
\newcommand{\POWHEGhvqHERWIG}{{\POWHEG \hvq}\,+\,{\HERWIG}\xspace}
\newcommand{\POWHEGbbfourlPYTHIA}{{\POWHEG \bbfourl}\,+\,{\PYTHIA}\xspace}
\newcommand{\bbfourl}{\ensuremath{\texttt{bb4l}}\xspace}
\newcommand{\MGvATNLO} {{\textsc{MG5\_aMC@NLO}}\xspace}

\newcommand{\tX}{\ensuremath{\PQt\mathrm{X}}\xspace}
\newcommand{\FOpQCDbackground}{{FO pQCD\,+\,back\-ground}\xspace}

\newcommand{\TeV}{\text{Te\kern-0.1em V}\xspace}
\newcommand{\GeV}{\text{Ge\kern-0.1em V}\xspace}

\newcommand{\Photo}{\includegraphics[height=35mm]{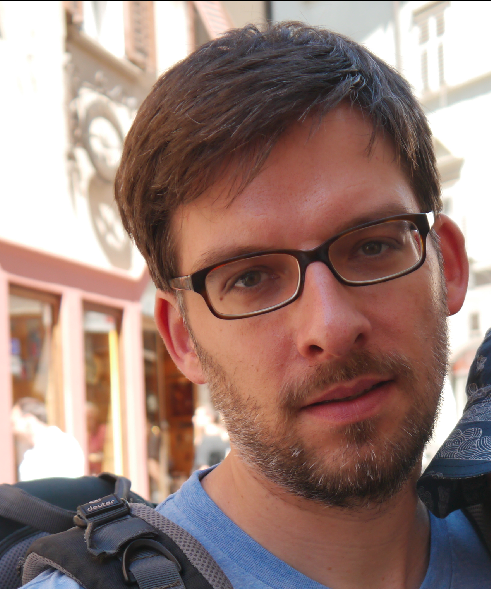}}

\begin{document}
\vspace*{4cm}
\title{Observation of a pseudoscalar excess at the \ttbar threshold}

\author{ A.~Grohsjean \\ (on behalf of the CMS Collaboration)}

\address{University of Hamburg, Faculty of Mathematics, Informatics and Natural Sciences, \\
Institute of Experimental Physics, Luruper Chaussee 149, 22761 Hamburg, Germany}

\maketitle\abstracts{
The CMS Collaboration recently observed a pseudoscalar excess of top quark–antiquark (\ttbar) events near the production threshold. One possible interpretation involves the formation of a short-lived quasi-bound state of \ttbar. This proceeding summarizes the first presentation of the results at the $59^{\textrm {th}}$ Rencontres de Moriond and outlines its potential implications.}

%\section{Introduction}

Thirty years after its discovery at the Fermilab Tevatron Collider,
%~\cite{CDF:1995wbb,D0:1995jca}
the top quark continues to play a central role in particle physics. As the most massive known elementary particle,
%~\cite{ATLAS:2024dxp}
it decays before hadronization. This allows its spin information to be transferred to its decay products, making top quark-antiquark (\ttbar) spin correlations a sensitive probe of the underlying production mechanism~\cite{CMS:2025kzt} and a powerful tool in the search for new physics~\cite{CMS:2025dzq}.
%~\cite{Bigi:1986jk,Bernreuther:2008ju}.

Although \ttbar pairs cannot form stable bound states due to the extremely short lifetime of the top quark, nonrelativistic quantum chromodynamics (NRQCD) predicts bound state enhancements at the production threshold~\cite{Fadin:1990wx,Kiyo:2008bv,Sumino:2010bv,Ju:2020otc,Fuks:2021xje,Garzelli:2024uhe}. The dominant contribution in proton-proton collisions comes from gluon-gluon fusion yielding a ${}^1S_0[1]$ state known as \etat. Other states, such as ${}^3P_0 [1]$ (\chit), are suppressed by powers of the top quark velocity, while color-octet configurations are suppressed below the \ttbar threshold because of a repulsive interaction between the top quarks, contributing significantly only above. 

To simulate contributions from \etat, we use a simplified model of a generic CP-odd, spin-0 color-singlet particle with direct couplings to gluons and top quarks~\cite{Maltoni:2024tul} and set its mass and width to match NRQCD predictions: $m_{\eta_t} = 343$\,GeV and $\Gamma_{\eta_t} = 2.8$\,GeV.

%\section{Event Selection and Observables}

The analysis targets final states with two oppositely charged leptons (electrons and/or muons) and at least two jets. Candidate events are required to contain exactly two leptons with transverse momentum $p_{\mathrm{T}} > 20$\,GeV, with at least one lepton satisfying $p_{\mathrm{T}} > 25$\,GeV. To suppress contributions from low-mass resonances, the invariant mass of the dilepton system must exceed 20\,GeV. In the $ee$ and $\mu\mu$ channels, backgrounds from $Z/\gamma^*$ processes are reduced by requiring the missing transverse momentum to be greater than 40\,GeV and the dilepton invariant mass to be at lest 15\,GeV different from the $Z$ boson mass. The presence of at least two jets with $p_{\mathrm{T}}>30$\,GeV and of at least one $b$-tagged jet with $p_{\mathrm{T}}>20$\,GeV is required, resulting in two or more jets per selected event. 
Events are reconstructed using a kinematic reconstruction algorithm~\cite{CMS-TOP-12-028}, which employs constraints from the $W$ boson and top quark masses, as well as the missing transverse momentum. The resolution of the reconstructed \ttbar invariant mass, \mttbar,  is approximately 15\% near the threshold and degrades to about 25\% at higher masses.
In addition to \mttbar, two angular observables are employed to probe \ttbar spin correlations. The variable \chel~\cite{Bernreuther:2015yna} is defined as the scalar product of the unit vectors of the momenta of the two leptons in the rest frames of their parent top quarks. The observable \chan is similarly constructed but includes a sign flip in the component parallel to the top quark direction. Both observables allow to discriminate scalar from pseudoscalar \ttbar events. 

%\section{Background Simulation}

The dominant background from nonresonant \ttbar production is simulated at next-to-leading order (NLO) in QCD using \POWHEGvtwohvqPYTHIA8.240. The sample is normalized to the inclusive cross section of $833.9^{+20.5}_{-30.0}$\,pb. 
To improve the modeling of differential distributions, the events are reweighted to predictions at NNLO in QCD with NLO electroweak corrections based on the double-differential distribution of \mttbar and the scattering angle of the top quark.
Alternative \ttbar samples are used for systematic studies: (i) \POWHEG matched to \HERWIG 7.2.2 with angular-ordered showering; (ii) \MGvATNLO with up to two additional jets at NLO; 
(iii) \bbfourl in \POWHEG vRES, simulating the full $\mathrm{pp} \to b\bar{b} \ell^+ \ell^- \nu \bar{\nu}$ process at NLO. The same NNLO QCD and NLO EW corrections are applied to these samples for consistency.
Minor reducible background (BG) contributions in the analysis originate from single top quark production (\tX), Drell-Yan production, diboson production, and \ttbar production in association with a vector boson.

\begin{figure*}[!t]
\centering
\includegraphics[width=0.32\textwidth]{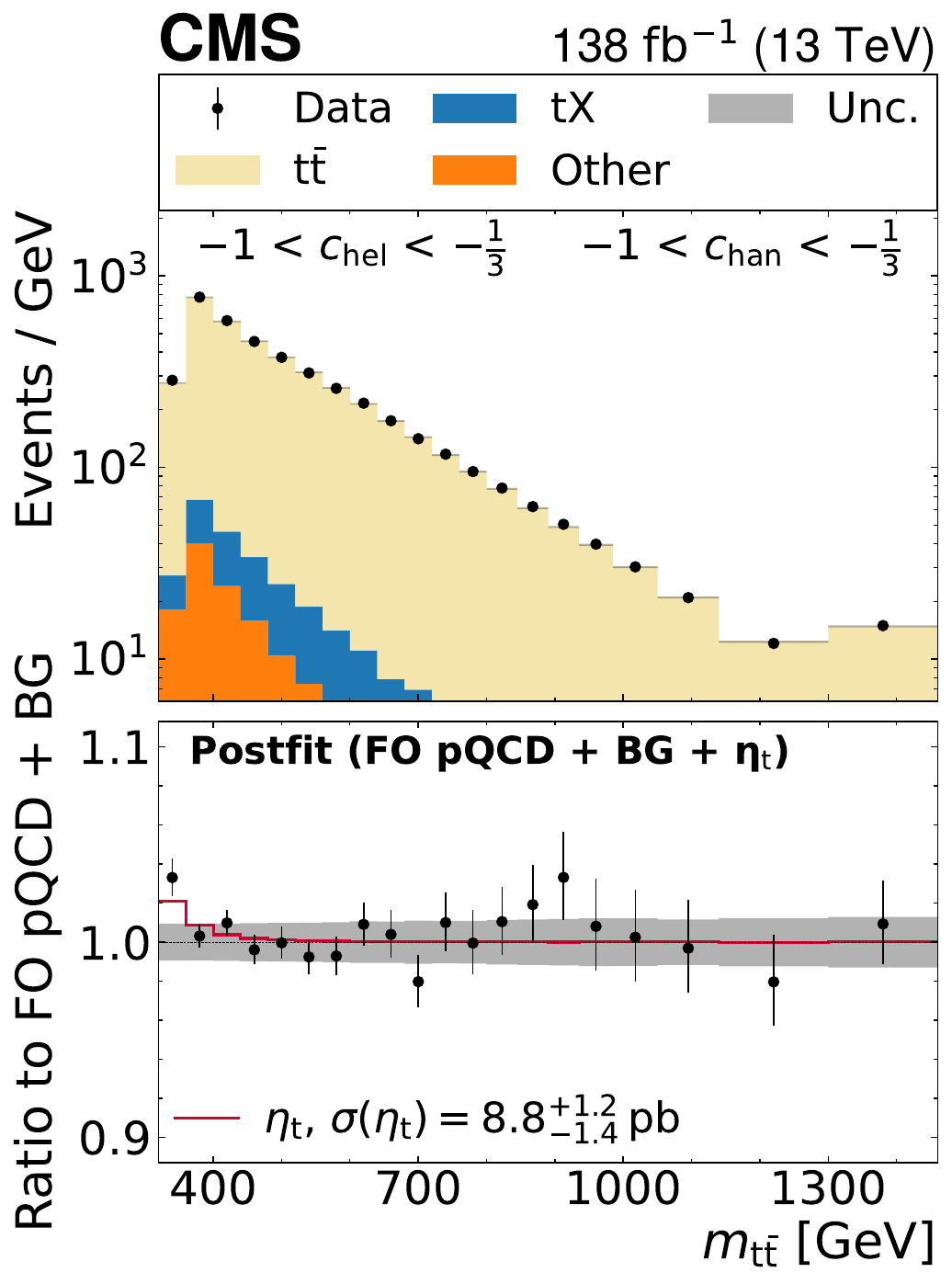}%
\hfill%
\includegraphics[width=0.32\textwidth]{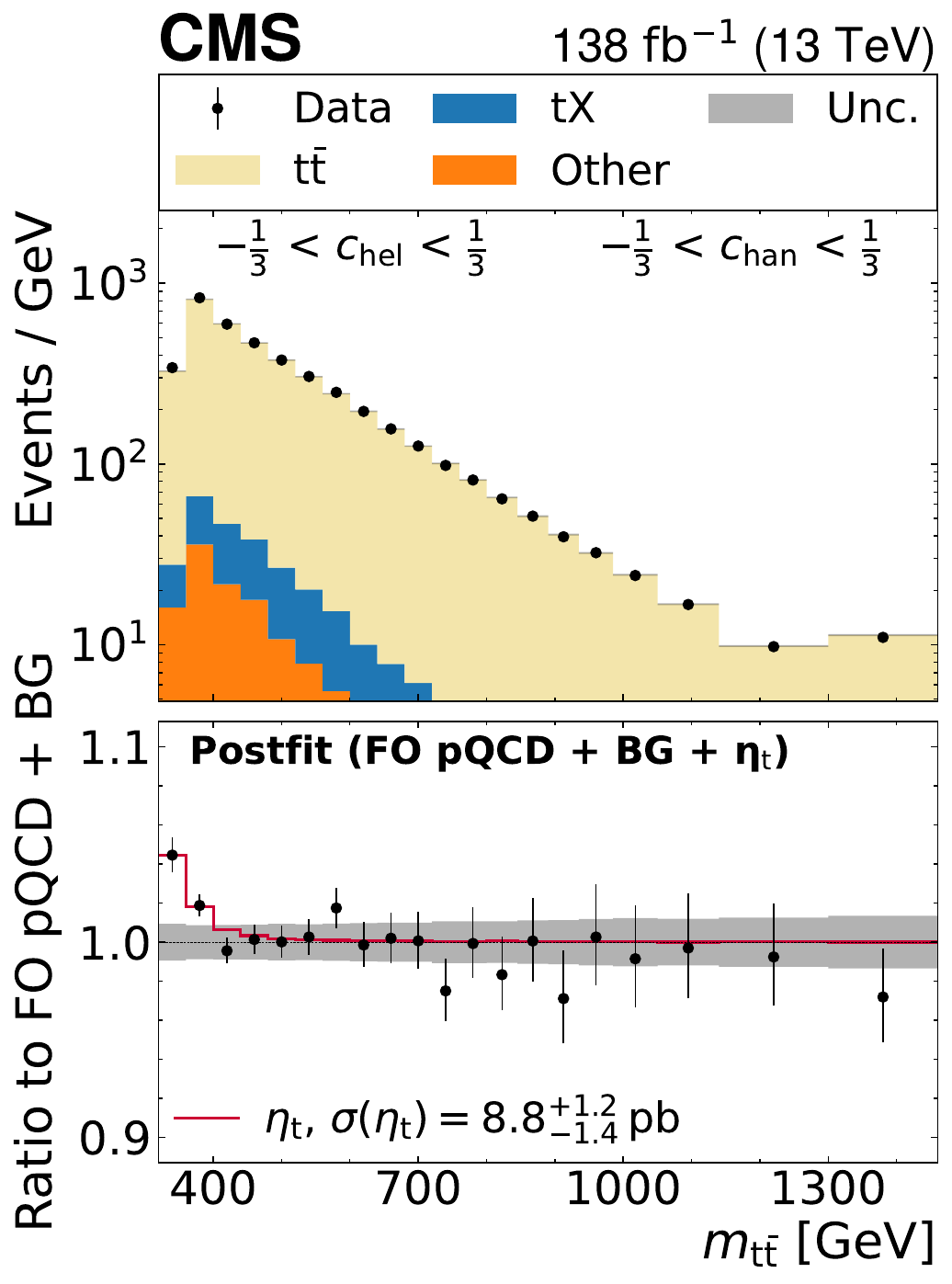}%
\hfill%
\includegraphics[width=0.32\textwidth]{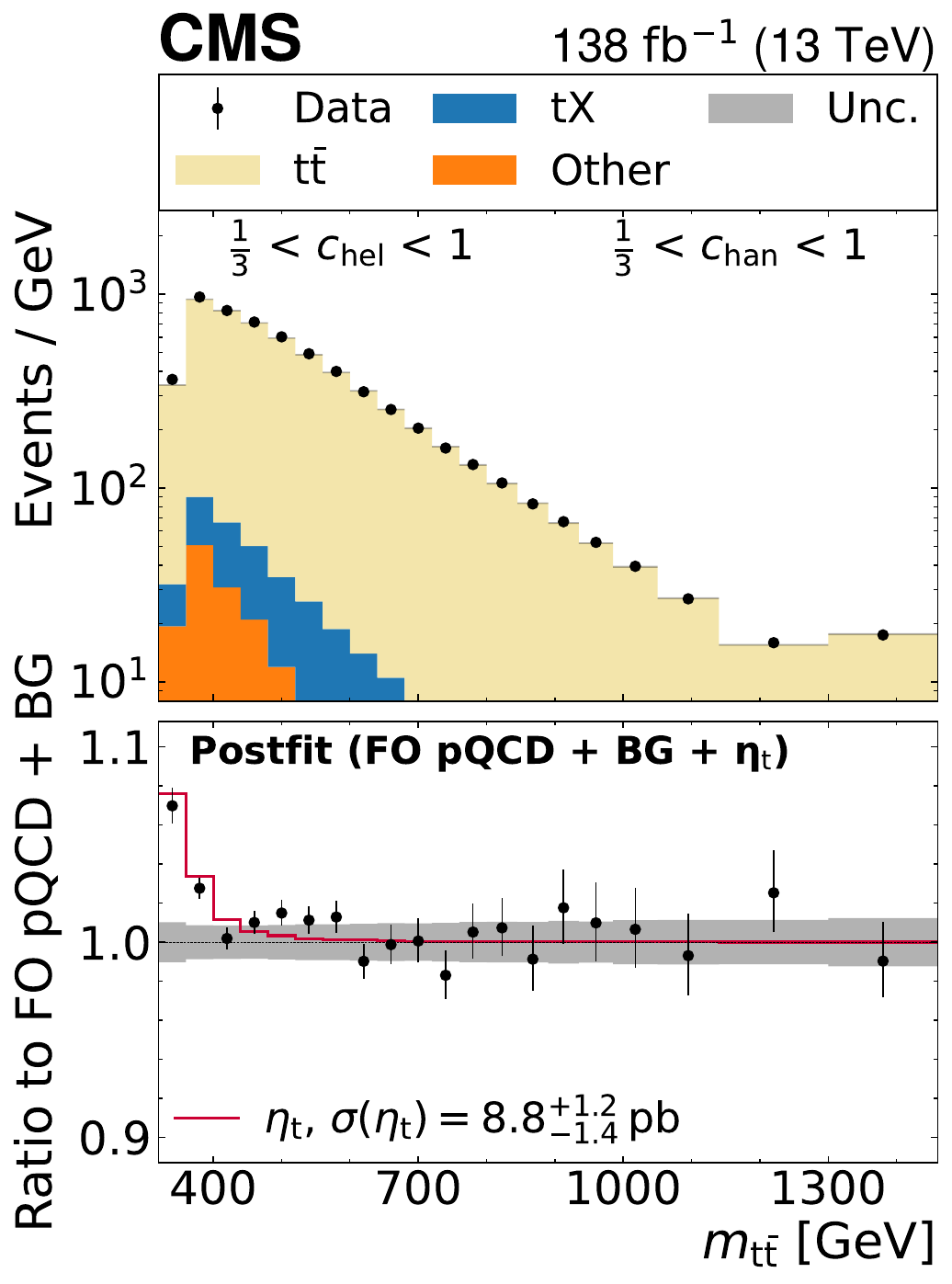}
\caption{
    Observed (points with statistical error bars) and predicted (stacked colored histograms) \mttbar distribution in three out of nine (\chel, \chan) bins.
    In the upper panels, the \ttbar histogram shows the FO pQCD prediction after the fit to the data. The shown event rates are divided by the bin width.
    The lower panels display the ratio of the data to the \FOpQCDbackground prediction, with the \etat signal overlaid at its fitted cross section (red line). The gray band indicates the postfit uncertainty.}
\label{fig:combined_ll}
\end{figure*}

%\section{Results}
A binned profile-likelihood fit is performed using the joint distribution of the reconstructed $m_{t\bar{t}}$, \chel, and \chan, discretized into $20{\times}3{\times}3$ bins. The analysis is interpreted in terms of a simplified model of $\eta_t$. The extracted production cross section is $\sigma(\eta_t) = 8.8 \pm 0.5~\text{(stat)}^{+1.1}_{-1.3}~\text{(syst)}$\,pb. Figure~\ref{fig:combined_ll} illustrates the observed excess in three representative bins of the angular observables, where the contribution from the $\eta_t$ signal is expected to be most pronounced in the most right panel. The measured cross section is compatible with a recent theory estimate of 6.4\,\unit{pb}~\cite{Fuks:2021xje}.

\begin{figure}[!htp]
\centering
%\hfill
\includegraphics[width=0.44\textwidth]{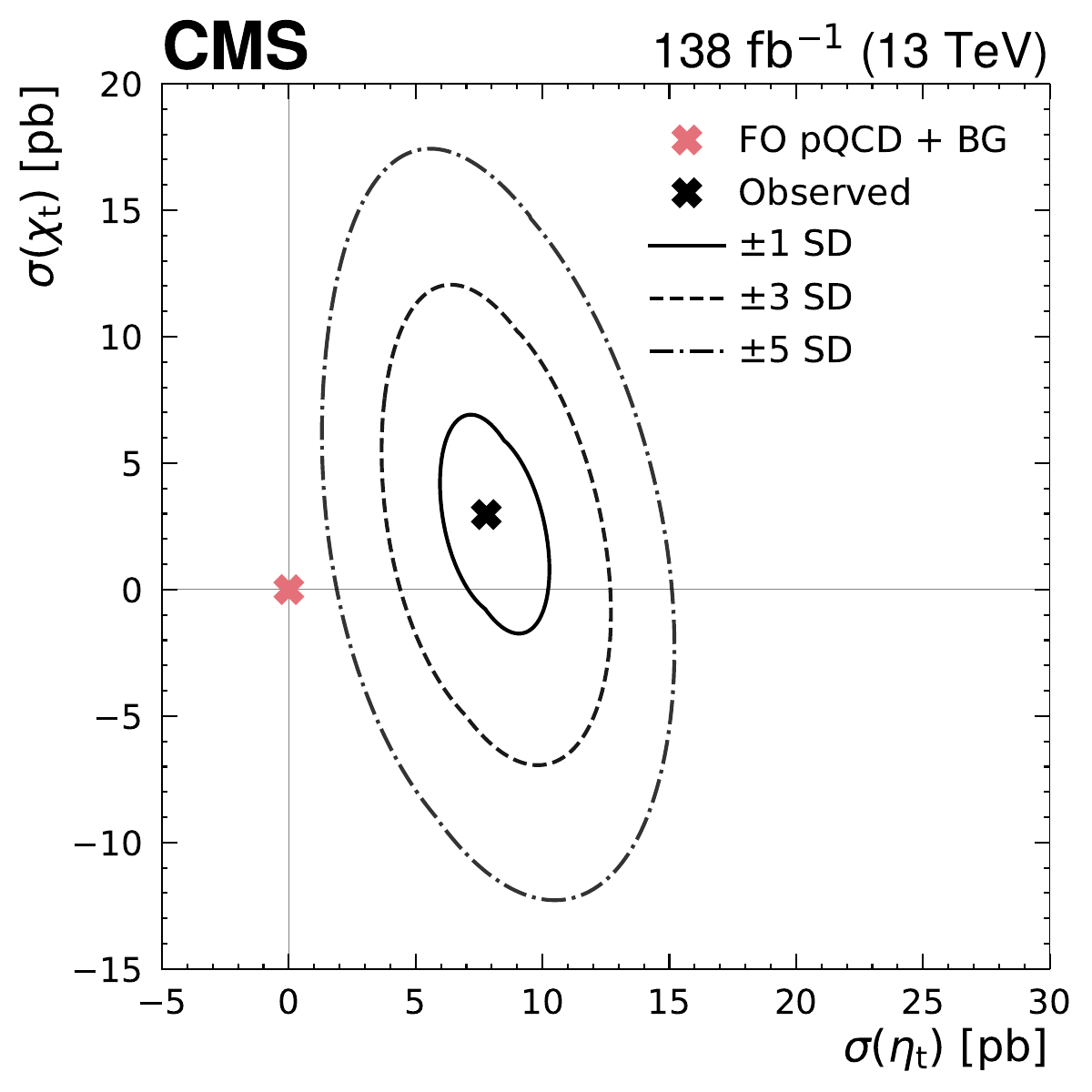}%
%\hfill
%\includegraphics[width=0.32\textwidth]{figures/Figure_002-b.pdf}
%\hfill
\caption{%
    Best fit value (cross) and allowed regions at one (solid line), three (dashed line), and five (dotted-dashed line) standard deviations for the cross section of \etat and \chit production, as observed in data (black). The \FOpQCDbackground expectation of zero \etat and \chit contributions is denoted by a red star.
    Negative cross section values refer to a reduction of the \ttbar cross sections relative to the FO pQCD prediction around the threshold.
}
\label{fig:res_a_vs_h_contour}
\end{figure}

To probe the spin and CP properties of the excess, a second fit is performed allowing simultaneous contributions from a pseudoscalar \etat and a scalar \chit state. The results shown in Fig.~\ref{fig:res_a_vs_h_contour} indicate no significant contribution from \chit, while the \etat component is favored with a significance exceeding five standard deviations, establishing the pseudoscalar nature of the excess.
The leading systematic uncertainties in the analysis are associated with the modeling of \ttbar production. The dominant contribution arises from comparing the nominal simulation to the \bbfourl prediction, which includes full off-shell effects and interference between \ttbar and single top. The next most significant uncertainty is related to parton shower modeling, particularly variations in the strong coupling constant for final-state radiation. Additional important sources include variations of the top quark mass and the Yukawa coupling, as well as the use of \HERWIG as an alternative parton shower model. 
Background-only fits fail to account for the observed discrepancies in both \mttbar and \chel, accompanied by strong pulls in nuisance parameters related to the NLO electroweak corrections applied to FO pQCD \ttbar production, either through the value of the top quark Yukawa coupling or the difference between the multiplicative and additive application scheme. 
\begin{figure*}[!htp]
\centering
\includegraphics[width=0.32\textwidth]{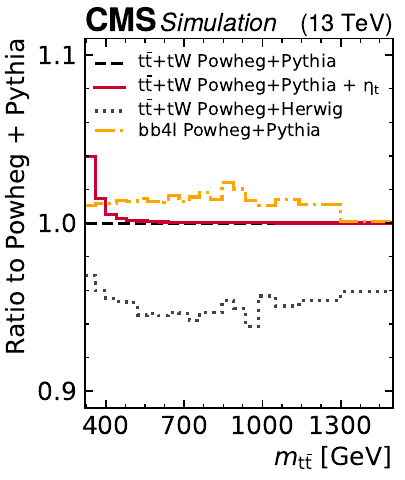}%
\hfill%
\includegraphics[width=0.32\textwidth]{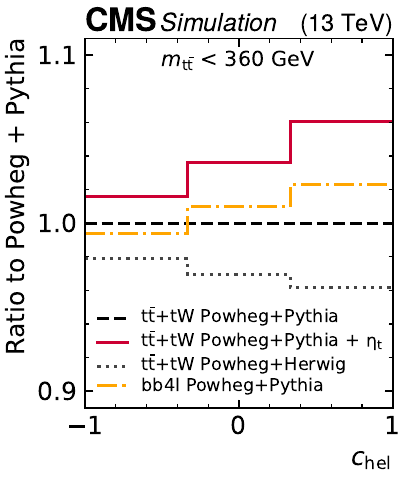}%
\hfill%
\includegraphics[width=0.32\textwidth]{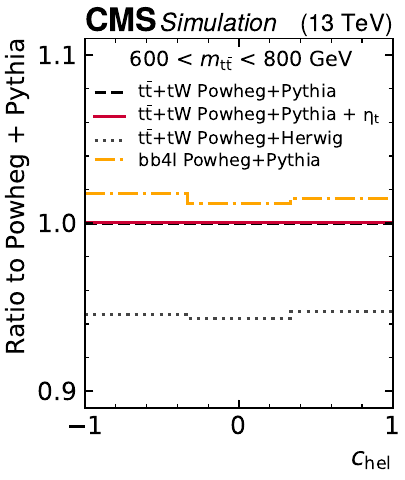}%
\caption{
    Ratios of the predictions for \POWHEGhvqHERWIG (gray), \POWHEGbbfourlPYTHIA (orange) to \POWHEGhvqPYTHIA (black) for \mttbar (left), \chel at the \ttbar threshold (center), and \chel in the \ttbar continuum (right), both integrated over \chan.
    The effect of adding \etat to \POWHEGhvqPYTHIA is shown in red for comparison. 
    }
    \label{fig:generators}
\end{figure*}
To illustrate the sensitivity to different generator setups, predictions from \POWHEGhvqHERWIG and \POWHEGbbfourlPYTHIA are compared in Fig.~\ref{fig:generators} to the nominal \POWHEGhvqPYTHIA configuration. \POWHEGhvqHERWIG predicts a lower overall \ttbar acceptance, an enhancement at low \mttbar values, and a reduced slope in the \chel distribution near the threshold - features that partially mimic or oppose the \etat signal and thus leave the extracted cross section largely unaffected. In contrast, the \bbfourl setup predicts a steeper \chel slope below threshold, similar to the \etat signal, leading to a reduced extracted cross section of $6.6 \pm 1.4$\,pb. Above the threshold, all generators yield consistent \chel shapes, as shown in Fig.~\ref{fig:generators} (right).
The observed excess is consistent with enhancements at the \ttbar threshold in previous differential cross section measurements reported by ATLAS~\cite{ATLAS:2023gsl} and CMS~\cite{CMS:2019esx}.
Similarly, both ATLAS~\cite{ATLAS:2023fsd} and CMS~\cite{CMS:2024pts} reported a mild tension between the observed and expected measurement of spin correlation in the \ttbar threshold region, which was used to study quantum entanglement. 
%A significant excess is observed in data at low \mttbar, consistent with the expectations from a simplified model of a pseudoscalar top quark–antiquark bound state, \etat. The excess exceeds five standard deviations in significance and cannot be attributed to known systematic uncertainties or alternative background predictions. 
Looking ahead, further improvements in background modeling, especially \ttbar off-shell production at NNLO and theoretical input on the \etat signal modeling, will be essential. Upcoming measurements in the semileptonic channel and potential confirmation by ATLAS will be key milestones. This intriguing result at the \ttbar threshold marks a potentially exciting new chapter in top quark physics.
\section*{Acknowledgments}
I acknowledge support by the Deutsche Forschungsgemeinschaft (DFG, German Research Foundation) under Germany‘s Excellence Strategy – EXC 2121 “Quantum Universe” – 390833306.
\section*{References}
\bibliography{grohsjean}

@article{CMS:2025kzt,
    author = "Hayrapetyan, A. and others",
    collaboration = "CMS",
    title = "{Observation of a pseudoscalar excess at the top quark pair production threshold}",
    eprint = "2503.22382",
    archivePrefix = "arXiv",
    primaryClass = "hep-ex",
    reportNumber = "CMS-TOP-24-007, CERN-EP-2025-061",
    doi = "10.1088/1361-6633/adf7d3",
    journal = "Rept. Prog. Phys.",
    volume = "88",
    pages = "087801",
    year = "2025"
}

@article{CMS:2025dzq,
    author = "Hayrapetyan, A. and others",
    collaboration = "CMS, EMAIL:cms-publication-committee-chair@cern.ch",
    title = "{Search for heavy pseudoscalar and scalar bosons decaying to a top quark pair in pp collisions at $\sqrt{s} = 13\,\textrm{TeV}$}",
    eprint = "2507.05119",
    archivePrefix = "arXiv",
    primaryClass = "hep-ex",
    reportNumber = "CMS-HIG-22-013, CERN-EP-2025-124",
    doi = "10.1088/1361-6633/ae2207",
    journal = "Rept. Prog. Phys.",
    volume = "88",
    pages = "127801",
    year = "2025"
}

@article{Fadin:1990wx,
    author = "Fadin, V. S. and Khoze, V. A. and Sj{\"o}strand, T.",
    title = "On the threshold behaviour of heavy top production",
    doi = "10.1007/BF01614696",
    journal = "Z. Phys. C",
    volume = "48",
    pages = "613",
    year = "1990"
}

@article{Maltoni:2024tul,
    author = "Maltoni, F. and Severi, C. and Tentori, S. and Vryonidou, E.",
    title = "Quantum detection of new physics in top-quark pair production at the {LHC}",
    eprint = "2401.08751",
    archivePrefix = "arXiv",
    primaryClass = "hep-ph",
    doi = "10.1007/JHEP03(2024)099",
    journal = "JHEP",
    volume = "03",
    pages = "099",
    year = "2024"
}

@article{Bernreuther:2015yna,
    author = "Bernreuther, W. and Heisler, D. and Si, Z.-G.",
    title = "A set of top quark spin correlation and polarization observables for the {LHC}: Standard model predictions and new physics contributions",
    eprint = "1508.05271",
    archivePrefix = "arXiv",
    primaryClass = "hep-ph",
    doi = "10.1007/JHEP12(2015)026",
    journal = "JHEP",
    volume = "12",
    pages = "026",
    year = "2015"
}

@article{Garzelli:2024uhe,
    author = "Garzelli, M. V. and Limatola, G. and Moch, S.-O. and Steinhauser, M. and Zenaiev, O.",
    title = "Updated predictions for toponium production at the {LHC}",
    eprint = "2412.16685",
    archivePrefix = "arXiv",
    primaryClass = "hep-ph",
    doi = "10.1016/j.physletb.2025.139532",
    journal = "Phys. Lett. B",
    volume = "866",
    pages = "139532",
    year = "2025"
}

@article{Fuks:2021xje,
    author = "Fuks, B. and Hagiwara, K. and Ma, K. and Zheng, Y.-J.",
    title = "Signatures of toponium formation in {LHC} run 2 data",
    eprint = "2102.11281",
    archivePrefix = "arXiv",
    primaryClass = "hep-ph",
    doi = "10.1103/PhysRevD.104.034023",
    journal = "Phys. Rev. D",
    volume = "104",
    pages = "034023",
    year = "2021"
}

@article{Ju:2020otc,
    author = "Ju, W.-L. and Wang, G. and Wang, X. and Xu, X. and Xu, Y. and Yang, L. L.",
    title = "Top quark pair production near threshold: single/double distributions and mass determination",
    eprint = "2004.03088",
    archivePrefix = "arXiv",
    primaryClass = "hep-ph",
    doi = "10.1007/JHEP06(2020)158",
    journal = "JHEP",
    volume = "06",
    pages = "158",
    year = "2020"
}

@article{Kiyo:2008bv,
    author = "Kiyo, Y. and K{\"u}hn, J. H. and Moch, S. and Steinhauser, M. and Uwer, P.",
    title = "Top-quark pair production near threshold at {LHC}",
    eprint = "0812.0919",
    archivePrefix = "arXiv",
    primaryClass = "hep-ph",
    doi = "10.1140/epjc/s10052-009-0892-7",
    journal = "Eur. Phys. J. C",
    volume = "60",
    pages = "375",
    year = "2009"
}

@article{Sumino:2010bv,
    author = "Sumino, Y. and Yokoya, H.",
    title = "Bound-state effects on kinematical distributions of top quarks at hadron colliders",
    eprint = "1007.0075",
    archivePrefix = "arXiv",
    primaryClass = "hep-ph",
    doi = "10.1007/JHEP09(2010)034",
    journal = "JHEP",
    volume = "09",
    pages = "034",
    year = "2010",
    note = "[Erratum: 10.1007/JHEP06(2016)037]"
}

@article{CMS:2019esx,
    author = "Sirunyan, A. M. and others",
    collaboration = "CMS",
    title = "Measurement of \ttbar normalised multi-differential cross sections in ${\Pp\Pp}$ collisions at $\sqrt{s}={13\,\TeV}$, and simultaneous determination of the strong coupling strength, top quark pole mass, and parton distribution functions",
    eprint = "1904.05237",
    archivePrefix = "arXiv",
    primaryClass = "hep-ex",
    doi = "10.1140/epjc/s10052-020-7917-7",
    journal = "Eur. Phys. J. C",
    volume = "80",
    pages = "658",
    year = "2020"
}

@article{CMS:2024pts,
    author = "Hayrapetyan, A. and others",
    collaboration = "CMS",
    title = "Observation of quantum entanglement in top quark pair production in proton-proton collisions at $\sqrt{s}={13\,\TeV}$",
    eprint = "2406.03976",
    archivePrefix = "arXiv",
    primaryClass = "hep-ex",
    doi = "10.1088/1361-6633/ad7e4d",
    journal = "Rep. Prog. Phys.",
    volume = "87",
    pages = "117801",
    year = "2024"
}

@article{ATLAS:2023gsl,
    author = "Aad, G. and others",
    collaboration = "ATLAS",
    title = "Inclusive and differential cross sections for dilepton \ttbar production measured in $\sqrt{s}={13\,\TeV}$ ${\Pp\Pp}$ collisions with the {ATLAS} detector",
    eprint = "2303.15340",
    archivePrefix = "arXiv",
    primaryClass = "hep-ex",
    doi = "10.1007/JHEP07(2023)141",
    journal = "JHEP",
    volume = "07",
    pages = "141",
    year = "2023"
}

@article{ATLAS:2023fsd,
    author = "Aad, G. and others",
    collaboration = "ATLAS",
    title = "Observation of quantum entanglement with top quarks at the {ATLAS} detector",
    eprint = "2311.07288",
    archivePrefix = "arXiv",
    primaryClass = "hep-ex",
    doi = "10.1038/s41586-024-07824-z",
    journal = "Nature",
    volume = "633",
    pages = "542",
    year = "2024"
}

@article{CMS-TOP-12-028,
    author = "Khachatryan, V. and others",
    collaboration = "CMS",
    title = "Measurement of the differential cross section for top quark pair production in ${\Pp\Pp}$ collisions at $\sqrt{s} $ = {8\,\TeV}",
    eprint = "1505.04480",
    archivePrefix = "arXiv",
    primaryClass = "hep-ex",
    doi = "10.1140/epjc/s10052-015-3709-x",
    journal = "Eur. Phys. J. C",
    volume = "75",
    pages = "542",
    year = "2015"
}
\end{document}